\pdfoutput=1
\PassOptionsToPackage{colorlinks=true,linkcolor=[HTML]{E35B2F},citecolor=[HTML]{E35B2F},urlcolor=[HTML]{E35B2F}}{hyperref}
\documentclass[11pt, letterpaper, onecolumn, numbering]{googledeepmind}
\usepackage[authoryear, sort&compress, round]{natbib}
\usepackage{graphicx} % Required for inserting images
\usepackage{svg} % For SVG logo support
\usepackage{amsmath}
\usepackage{amssymb} % for \mathbb and additional math symbols
\usepackage{booktabs}
\usepackage{makecell}
\usepackage{enumitem}
\usepackage[table]{xcolor}
\definecolor{pastelorange}{RGB}{255, 235, 215}
\usepackage{tikz}
\hypersetup{
  colorlinks=true,
  linkcolor=[HTML]{E35B2F},
  citecolor=[HTML]{E35B2F},
  urlcolor=[HTML]{E35B2F}
}
\renewcommand{\today}{}
\let\cite\citep
\makeatletter
\let\@@par\par
\makeatother

\title{Paris: A Decentralized Trained Open-Weight Diffusion Model}
\author{Zhiying Jiang, Raihan Seraj, Marcos Villagra, Bidhan Roy\textsuperscript{*}\\
\vspace{0.2em}
{\small Bagel Labs $\cdot$ \href{https://bagel.com}{bagel.com}}}
\reportnumber{}

\AtBeginDocument{%
  \fancypagestyle{firststyle}{%
    \fancyhf{}%
    \fancyhead[L]{}%
    \fancyhead[C]{}%
    \fancyhead[R]{}%
    \fancyfoot[L]{\footerfont\textsuperscript{*}Corresponding author: bidhan[at]bagel[dot]com}%
    \fancyfoot[R]{\footerfont\thepage}%
    \fancyfoot[C]{}%
    \renewcommand{\headrulewidth}{0pt}%
    \renewcommand{\footrulewidth}{1pt}%
    \renewcommand{\footrule}{\textcolor{gray}{\vskip-\footruleskip\vskip-\footrulewidth\rule{\textwidth}{1pt}}}%
  }%
}

\newcommand{\projectlinks}{%
\noindent
\begin{minipage}{0.7\textwidth}
{\footnotesize
\href{https://github.com/bageldotcom/paris}{\raisebox{-0.1em}{\includegraphics[height=0.9em]{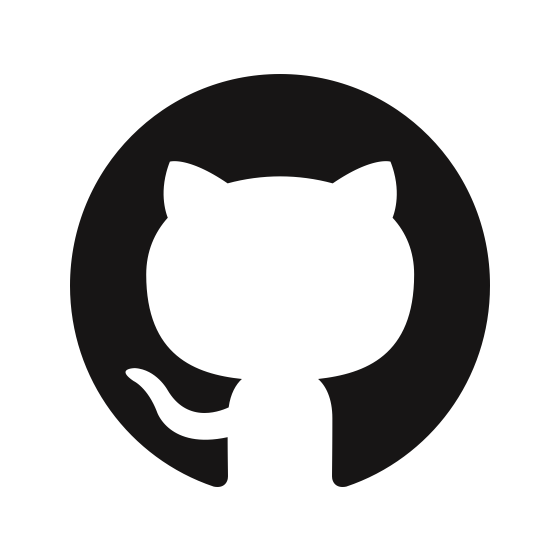}} \texttt{bageldotcom/paris}} \quad
\href{https://huggingface.co/bageldotcom/paris}{\raisebox{-0.1em}{\includegraphics[height=0.9em]{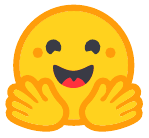}} \texttt{bageldotcom/paris}}
}
\end{minipage}%
\hfill
\begin{minipage}{0.25\textwidth}
\raggedleft
\includesvg[height=0.9em]{Vector.svg}
\end{minipage}
}

\newlength{\gridcardwidth}
\newlength{\gridcardheight}
\begin{abstract}
    We present \textbf{Paris}, the first publicly released diffusion model pre-trained entirely through decentralized computation. Paris demonstrates that high-quality text-to-image generation can be achieved without centrally coordinated infrastructure. \textbf{Paris is open for research and commercial use.}

    \vspace{0.5em}
    
    Paris required implementing our Distributed Diffusion Training framework from scratch. The model consists of 8 expert diffusion models (129M-605M parameters each) trained in complete isolation \textbf{with no gradient, parameter, or intermediate activation synchronization}. Rather than requiring synchronized gradient updates across thousands of GPUs, we partition data into semantically coherent clusters where each expert independently optimizes its subset while collectively approximating the full distribution. A lightweight transformer router dynamically selects appropriate experts at inference, achieving generation quality comparable to centrally coordinated baselines.
    
    \vspace{0.5em}

Eliminating synchronization enables training on heterogeneous hardware without specialized interconnects. Empirical validation (Section~\ref{sec:experiments}) confirms that Paris's decentralized training maintains generation quality while removing the dedicated GPU cluster requirement for large-scale diffusion models. Paris achieves this using 14$\times$ less training data and 16$\times$ fewer GPU-days than the prior decentralized baseline (Table~\ref{tab:comparison_sota}, Figure~\ref{fig:efficiency_plot}).
\end{abstract}

\begin{document}

% Remove horizontal rule on first page only
\makeatletter
\let\oldfirststyle\ps@firststyle
\def\ps@firststyle{%
  \oldfirststyle%
  \def\headrule{}%
}
\makeatother

% Add pastel orange background box with margins
\begin{tikzpicture}[remember picture,overlay]
\fill[pastelorange, rounded corners=8pt] ([xshift=1.5cm,yshift=-2.3cm]current page.north west) rectangle ([xshift=-1.5cm,yshift=-18cm]current page.north east);
\end{tikzpicture}

\maketitle

% Project links below abstract, above introduction
\vskip20pt
\projectlinks
\vspace{2.5em}
% Adjust grid to fit in 1/3 of page
\begin{figure}[h]
\centering
\begingroup
\setlength{\tabcolsep}{0.2em}
\renewcommand{\arraystretch}{0}
\resizebox{0.95\textwidth}{!}{%
\begin{tabular}{ccccc}
\includegraphics[width=\gridcardwidth,height=\gridcardheight,keepaspectratio=false]{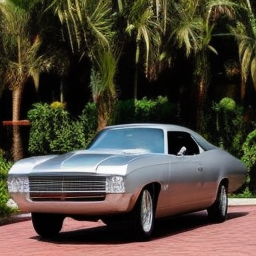} &
\includegraphics[width=\gridcardwidth,height=\gridcardheight,keepaspectratio=false]{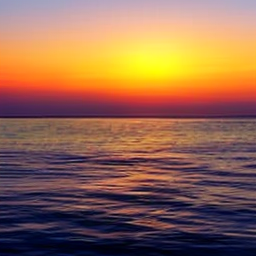} &
\includegraphics[width=\gridcardwidth,height=\gridcardheight,keepaspectratio=false]{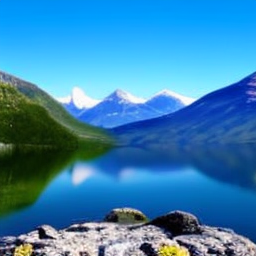} &
\includegraphics[width=\gridcardwidth,height=\gridcardheight,keepaspectratio=false]{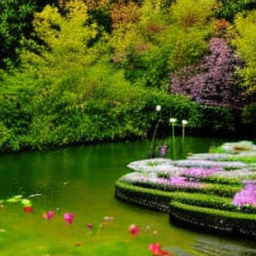} &
\includegraphics[width=\gridcardwidth,height=\gridcardheight,keepaspectratio=false]{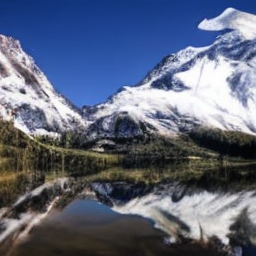} \\[2pt]
\parbox{\gridcardwidth}{\centering\fontsize{5}{6}\selectfont\color{black!65} A silver car on a brick road surrounded by palm trees} &
\parbox{\gridcardwidth}{\centering\fontsize{5}{6}\selectfont\color{black!65} Sunset over a calm ocean} &
\parbox{\gridcardwidth}{\centering\fontsize{5}{6}\selectfont\color{black!65} A crystal-clear alpine lake reflecting the majestic mountain} &
\parbox{\gridcardwidth}{\centering\fontsize{5}{6}\selectfont\color{black!65} A photograph of a majestic garden} &
\parbox{\gridcardwidth}{\centering\fontsize{5}{6}\selectfont\color{black!65} A beautiful landscape} \\[1em]
\includegraphics[width=\gridcardwidth,height=\gridcardheight,keepaspectratio=false]{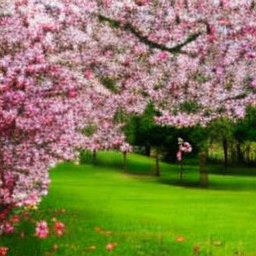} &
\includegraphics[width=\gridcardwidth,height=\gridcardheight,keepaspectratio=false]{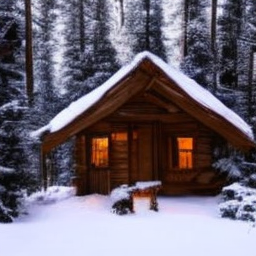} &
\includegraphics[width=\gridcardwidth,height=\gridcardheight,keepaspectratio=false]{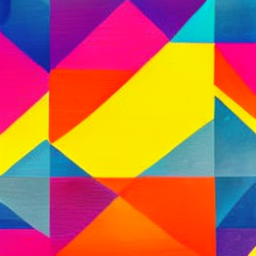} &
\includegraphics[width=\gridcardwidth,height=\gridcardheight,keepaspectratio=false]{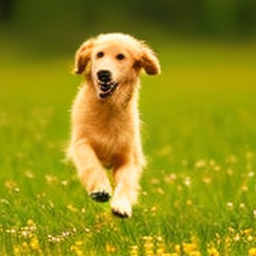} &
\includegraphics[width=\gridcardwidth,height=\gridcardheight,keepaspectratio=false]{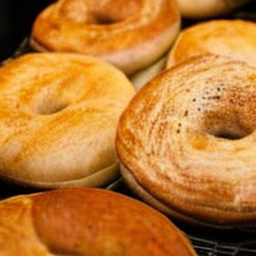} \\[2pt]
\parbox{\gridcardwidth}{\centering\fontsize{5}{6}\selectfont\color{black!65} A peaceful garden with blooming cherry trees} &
\parbox{\gridcardwidth}{\centering\fontsize{5}{6}\selectfont\color{black!65} A cozy cabin in a snowy forest} &
\parbox{\gridcardwidth}{\centering\fontsize{5}{6}\selectfont\color{black!65} Abstract art with vibrant colors and geometric shapes} &
\parbox{\gridcardwidth}{\centering\fontsize{5}{6}\selectfont\color{black!65} A high-resolution photograph of a golden retriever puppy running through a meadow} &
\parbox{\gridcardwidth}{\centering\fontsize{5}{6}\selectfont\color{black!65} A high-quality photograph of fresh bagels, out of the oven}

\end{tabular}%
}
\endgroup
\caption*{Text conditioned image generation samples using Paris}
\label{fig:paris_samples}
\end{figure}
\section{Introduction}

Large-scale diffusion model training requires synchronized gradient updates across thousands of GPUs connected through high-bandwidth interconnects. State-of-the-art text-to-image models like Stable Diffusion required 150,000 A100 GPU-hours~\citep{rombach2022high}, while Google's Imagen required hundreds of TPU-v4 chips~\citep{saharia2022photorealistic}, demanding dedicated clusters with InfiniBand or similar networking. This creates two fundamental barriers:
\begin{enumerate}[label=(\roman*)]
\item Only institutions with massive computing infrastructure can train these models.
\item The synchronization requirement prevents leveraging geographically distributed and/or commodity hardware lacking specialized interconnects.
\end{enumerate}

We present \textbf{Paris}, demonstrating that high-quality text-to-image generation can be achieved without centrally coordinated infrastructure. The model consists of 8 independently trained expert diffusion models that exchange neither gradients, parameters, nor intermediate activations during training. This complete computational independence required implementing our Distributed Diffusion Training framework from scratch, which extends the decentralized
flow matching theory of~\citet{mcallister2025decentralized}, the Diffusion Transformer architecture of~\citet{Peebles2022DiT}, and architectural optimizations from PixArt-$\alpha$~\citep{chen2023pixartalpha} to achieve practical, production-scale implementation.

Key insights enabling this approach include the natural decomposition of flow matching objectives across data partitions, allowing experts to optimize locally while the ensemble approximates the global distribution, the superior scaling properties of Diffusion Transformers for distributed training, and parameter-efficient conditioning that reduces model size without quality loss. These advances enable a lightweight routing mechanism learned post-hoc to orchestrate expert collaboration during inference without requiring any coordination during training.

This eliminates the need for specialized interconnects and dedicated clusters. Empirical validation confirms that fully decentralized training maintains generation quality while enabling development on fragmented compute resources previously unusable for large-scale diffusion models.

\textbf{Contributions.} Our key contributions are: (i) the first open-weight text-to-image diffusion model trained entirely through decentralized computation with zero inter-expert communication; (ii) extending DDM's framework~\citep{mcallister2025decentralized} with architectural and computational optimizations that achieve comparable generation quality while using only 1/14 of the training data (11M vs. 154M images) and 1/16 of the GPU-days; and (iii) complete open-source implementation demonstrating practical deployment on heterogeneous, geographically distributed hardware.

\section{Method}

\begin{figure*}[t]
    \centering
    \includegraphics[width=1\linewidth]{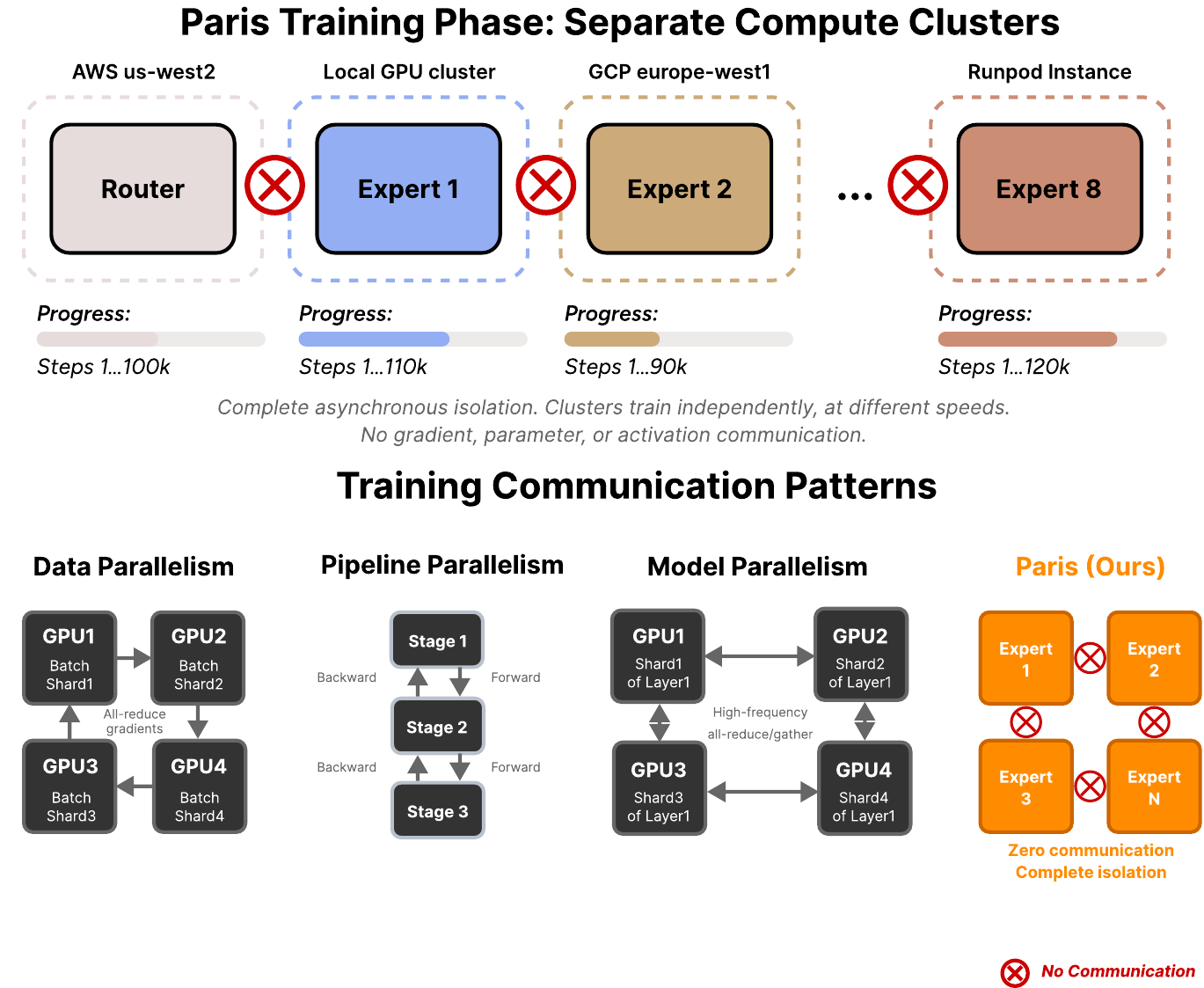}
        \caption{Multi-expert training pipeline of Paris.}
    \label{fig:paris}
\end{figure*}

\subsection{Overview}

We pretrain our model, Paris, utilizing our Distributed Diffusion Training framework, to enable fully decentralized training that doesn't require communication among experts. In order to achieve that, we partitioned the training set into $K$ clusters. We then train each expert on a partition \emph{in isolation}; and a lightweight router fuses experts at inference. 

Concretely, we: (i) pre-encode images into a latent space with a pretrained VAE~\citep{kingma2014auto} (``sd-vae-ft-mse'') to reduce compute, following~\citet{rombach2022high}; (ii) compute semantic features with DINOv2 and cluster the data; (iii) train $K$ expert denoisers; (iv) train a small transformer router \textit{independently}; and (v) perform different expert selection strategies at sampling.

\subsection{Implementation Scale and Architecture}

Paris extends decentralized flow matching theory~\citep{mcallister2025decentralized,lipman2023flow}, the Diffusion Transformer architecture~\citep{Peebles2022DiT}, and architectural optimizations from PixArt-$\alpha$~\citep{chen2023pixartalpha}. We implement decentralized training using Diffusion Transformers for their superior scaling properties compared to U-Nets~\citep{ronneberger2015u}, with initialization schemes ensuring convergence without coordination.

We validate Paris training recipe at two scales: \textbf{DiT-B/2} (where B denotes Base model size with 768 hidden dimensions, and /2 indicates patch size 2 for latent tokenization) with 129M parameters per expert and \textbf{DiT-XL/2} with 605M parameters per expert, totaling 1.03B and 4.84B parameters respectively across 8 experts. The framework partitions 11M LAION-Aesthetic~\citep{schuhmann2022laion} images into semantically coherent clusters using DINOv2 embeddings, enabling each expert to specialize on distinct visual domains like portraits, landscapes, or architecture.

Paris's routing mechanism operates directly on partially denoised latents throughout the reverse diffusion process, requiring the router to identify appropriate experts despite observing only noisy intermediate states. This noise-aware routing, combined with flow matching objectives using velocity prediction and initialization strategies, ensures stable convergence despite the complete absence of inter-expert coordination during training.

\subsection{Decentralized Flow Matching Objective}

Following the works of~\citet{mcallister2025decentralized} and~\citet{lipman2023flow}, we decompose the standard flow matching objective across $K$ expert models, each trained on disjoint data partitions. The marginal flow $u_t(x_t)$ transports samples from the noise distribution to the data distribution through:

\begin{equation}
u_t(x_t) = \int_{x_0} u_t(x_t|x_0) \frac{p_t(x_t|x_0)q(x_0)}{p_t(x_t)} dx_0,
\end{equation}
where $x_t = \alpha_t x_0 + \sigma_t \epsilon$ represents the noisy latent at timestep $t \in [0,1]$, with $\alpha_t = 1 - t$ and $\sigma_t = t$ for linear scheduling. In the discrete case over dataset $\mathcal{D}$:

\begin{equation}
u_t(x_t) = \frac{1}{p_t(x_t)} \sum_{x_0 \in \mathcal{D}} u_t(x_t|x_0)p_t(x_t|x_0)q(x_0).
\end{equation}
The key insight of DDM is partitioning the data into $K$ disjoint clusters $\{S_1, S_2, \ldots, S_K\}$, enabling decomposition:

\begin{equation}
u_t(x_t) = \sum_{k=1}^{K} p_t(k|x_t) \cdot u_t^{(k)}(x_t),
\end{equation}
where $u_t^{(k)}(x_t)$ is the flow predicted by expert $k$ trained only on cluster $S_k$, and $p_t(k|x_t)$ is the posterior probability from the router network. This formulation allows each expert to optimize independently:

\begin{equation}
\mathcal{L}_{\text{expert}}^{(k)} = \mathbb{E}_{x_0 \in S_k, t} \left[ \|v_{\theta_k}(x_t, t) - (x_0 - x_t)\|^2 \right],
\end{equation}
where $\theta_k$ denotes the parameters of expert $k$, $v_{\theta_k}(x_t, t)$ is the predicted velocity field at noisy latent $x_t$ and timestep $t \sim \mathcal{U}[0,1]$, $(x_0 - x_t)$ represents the target velocity direction from the noisy state back to the clean data, and the expectation is taken over clean samples $x_0$ drawn from cluster $S_k$ and uniformly sampled timesteps.

\subsection{Expert Models}

The DiTExpert architecture represents our adaptation of the Diffusion Transformer (DiT) framework for decentralized diffusion modeling, incorporating architectural innovations from both the original DiT paper and recent advances in text-to-image generation such as PixArt-$\alpha$. This architecture serves as the backbone for each expert model in our decentralized ensemble, providing superior scalability compared to traditional U-Net~\citep{ronneberger2015u} architectures while maintaining computational efficiency through strategic design choices.

\subsubsection{Core Architecture}

At its foundation, DiTExpert operates on latent representations rather than raw pixels,
following the latent diffusion paradigm of~\citet{rombach2022high}. 
The model processes $32 \times 32$ latent tensors with 4 channels, obtained through a pre-trained VAE encoder that performs $8\times$ spatial downsampling from the original $256 \times 256$ images. 
Following the Vision Transformer paradigm of~\citet{dosovitskiy2020image}, we incorporate fixed sinusoidal positional embeddings~\citep{vaswani2017attention} to preserve spatial relationships between patches.
Temporal information, critical for the diffusion process, is injected through a specialized timestep embedder.

\subsubsection{Transformer Blocks with Adaptive Layer Normalization}

The core of DiTExpert consists of various transformer blocks, each implementing a modified architecture that integrates timestep conditioning through Adaptive Layer Normalization (AdaLN). Unlike standard transformers that use fixed layer normalization~\citep{ba2016layer}, AdaLN modulates the normalized activations based on the timestep embedding:

\begin{equation}
\text{AdaLN}(h, c) = \gamma_c \odot \text{LayerNorm}(h) + \beta_c,
\end{equation}
where $\odot$ denotes the Hadamard product, $\gamma_c, \beta_c \in \mathbb{R}^D$ are scale and shift parameters predicted from the conditioning signal $c = \tau(t)$ through block-specific MLPs. This mechanism allows the model to adaptively adjust its behavior across different noise levels without increasing the sequence length.

Each DiT block follows the structure:
\begin{align}
h' &= h + \text{MSA}(\text{AdaLN}(h, c)), \\
h'' &= h' + \text{CrossAttn}(\text{AdaLN}(h', c), e_{\text{text}}) \quad \text{(if text-conditioned)}, \\
h_{\text{out}} &= h'' + \text{FFN}(\text{AdaLN}(h'', c)),
\end{align}
where MSA denotes multi-head self-attention~\citep{vaswani2017attention} with multiple heads, CrossAttn represents optional cross-attention for text conditioning, and FFN is a feed-forward network with expansion ratio 4.0 and GELU~\citep{hendrycks2016gaussian} activation. For text-conditional generation, we extend the base DiT architecture with cross-attention layers positioned between self-attention and feed-forward blocks. Text embeddings from frozen CLIP~\citep{radford2021learning} encoders are projected to the model's hidden dimension through a learned linear transformation.

\subsubsection{Parameter-Efficient AdaLN-Single Variant}

We implement an optional AdaLN-Single variant proposed by PixArt-$\alpha$. This modification computes conditioning signals once globally and uses learnable per-block embeddings:

\begin{equation}
[\gamma_1, \beta_1, ..., \gamma_L, \beta_L] = \text{MLP}_{\text{global}}(\tau(t)) + E_{\text{blocks}},
\end{equation}
where $E_{\text{blocks}} \in \mathbb{R}^{L \times 6D}$ are learned embeddings for $L$ blocks, each containing 6 modulation parameters, which helps reducing parameters substantially.

\begin{figure*}[t]
    \centering
    \includegraphics[width=1\linewidth]{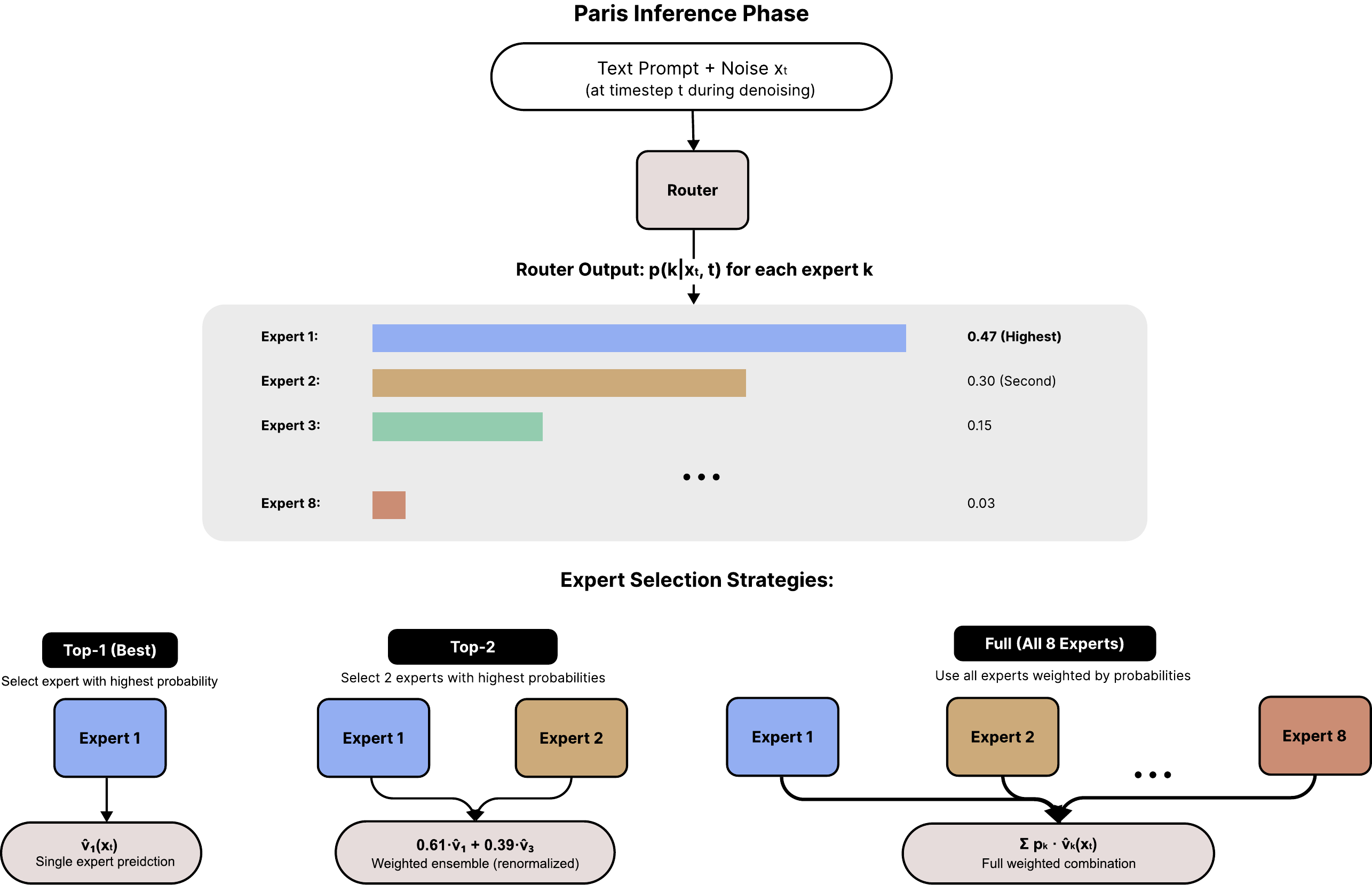}
        \caption{Multi-expert inference pipeline of Paris.}
    \label{fig:paris_inference}
\end{figure*}

\subsection{Router Networks}

The DiTRouter architecture represents a critical innovation in our decentralized diffusion framework, employing a lightweight Diffusion Transformer variant specifically designed for dynamic expert selection. Unlike traditional routing mechanisms that operate on clean inputs or require separate encoders, DiTRouter directly processes noisy latents at arbitrary timesteps, learning to identify which expert's training distribution best matches the current sample. This design enables coherent expert collaboration during the reverse diffusion process while maintaining computational efficiency.

\subsubsection{Core Architectural}

DiTRouter adopts the DiT-B/2 configuration, utilizing a smaller transformer architecture compared to the expert models to minimize routing overhead. The router processes the same $32 \times 32 \times 4$ latent representations as the experts but with reduced capacity, occupying around one-third the size of each expert model.

The fundamental task of the router is formulated as a classification problem over noisy inputs:

\begin{equation}
p(k|x_t, t) = \text{Router}_\phi(x_t, t),
\end{equation}
where $k \in \{1, ..., K\}$ indexes the experts, $x_t$ is the noisy latent at timestep $t$. This probabilistic formulation enables both hard routing (top-1 selection) and soft ensemble strategies during inference.

\subsubsection{Timestep-Aware Processing}

A key insight in router design is that expert specialization may vary across noise levels, certain experts might excel at high-frequency detail recovery (low noise) while others specialize in global structure formation (high noise). DiTRouter incorporates timestep information through the same sinusoidal embedding and MLP projection as the experts:

\begin{equation}
\tau_{\text{router}}(t) = \text{MLP}_{\text{router}}(\text{SinusoidalEmbed}(t)).
\end{equation}

This temporal conditioning is crucial for maintaining consistency with expert models, as both router and experts must interpret timesteps identically to ensure coherent collaboration. The router learns to adapt its classification strategy based on the denoising stage, potentially routing to different experts at different points in the generation process.

\subsubsection{Training Objective and Loss Function}

DiTRouter training employs a straightforward cross-entropy loss against ground-truth cluster assignments:

\begin{equation}
\mathcal{L}_{\text{router}} = \mathbb{E}_{x_0 \sim \mathcal{D}, t \sim [0,1]} \left[ -\log p_\phi(c(x_0) | x_t, t) \right],
\end{equation}
where $c(x_0) \in \{1, ..., K\}$ denotes the cluster assignment of the clean sample $x_0$. Critically, the router must learn to identify the correct expert despite only observing noisy versions of the input, requiring it to develop robust features invariant to diffusion noise while remaining sensitive to semantic content.

During training, we sample timesteps uniformly and apply the corresponding noise schedule to create $x_t$, ensuring the router gains experience across all noise levels. This comprehensive training enables accurate routing throughout the entire reverse diffusion process.

\subsection{DINOv2-Based Semantic Clustering}

We employ DINOv2~\citep{oquab2023dinov2}, a self-supervised vision transformer trained on 142M images, to extract semantically meaningful features that guide expert specialization through data partitioning. For each image, we extract 1024-dimensional features using DINOv2-ViT-L/14 on center-cropped $224 \times 224$ images: $f_i = \text{AvgPool}(\text{DINOv2}(x_i)) \in \mathbb{R}^{1024}$. 
To handle large-scale datasets efficiently, we adopt a two-stage hierarchical clustering strategy following~\citet{mcallister2025decentralized}: first performing fine-grained k-means~\cite{macqueen1967some} clustering to identify fine-grained initial centroids, then consolidating these into $K$ coarse clusters (typically $K=8$) through a second round of clustering.

\subsection{Inference Strategies for Decentralized Generation}

At inference time, the decentralized diffusion framework offers multiple strategies for combining expert predictions, each presenting distinct trade-offs between generation quality, computational cost, and sample diversity. As shown in Figure~\ref{fig:paris_inference}, these strategies leverage the router's learned expertise to dynamically orchestrate expert collaboration throughout the reverse diffusion process.

\subsubsection{Top-1 Expert Selection}

The most computationally efficient strategy routes each denoising step to a single expert based on the router's highest-confidence prediction:

\begin{equation}
v_t(x_t) = v_{\theta_{k^*}}(x_t, t), \quad k^* = \arg\max_k p_\phi(k|x_t, t).
\end{equation}

This hard routing approach maintains constant computational cost regardless of ensemble size, requiring only one expert forward pass per denoising step. Surprisingly, this simple strategy often yields the best results according to~\citet{mcallister2025decentralized}. While our experiments on smaller networks (Table~\ref{tab:fid_mono_multi_laion}) show a different pattern, Top-1 routing achieves competitive quality and Top-2 surpasses both monolithic and full ensemble approaches.

\subsubsection{Top-K Weighted Ensemble}

The Top-K strategy combines predictions from the $K'$ most relevant experts, weighted by their renormalized router probabilities:

\begin{equation}
v_t(x_t) = \sum_{k \in \text{Top-K}} \frac{p_\phi(k|x_t, t)}{\sum_{j \in \text{Top-K}} p_\phi(j|x_t, t)} \cdot v_{\theta_k}(x_t, t).
\end{equation}

This approach provides a middle ground between computational efficiency and ensemble benefits. The renormalization ensures the weights sum to unity, maintaining proper velocity scaling in the flow matching framework.

The implementation supports per-sample Top-K selection, allowing different samples in a batch to utilize different expert combinations based on their individual characteristics. This flexibility proves particularly valuable when generating diverse batches where samples may benefit from different specializations.

\subsubsection{Full Ensemble Integration}

The full ensemble strategy incorporates all experts weighted by their router probabilities:

\begin{equation}
v_t(x_t) = \sum_{k=1}^{K} p_\phi(k|x_t, t) \cdot v_{\theta_k}(x_t, t).
\end{equation}

While computationally expensive, requiring $K$ forward passes per step, this strategy theoretically provides the most accurate approximation to the true data distribution by fully utilizing the learned decomposition. However, empirical results often show diminishing returns or even degraded quality compared to selective strategies, potentially due to interference from less-relevant experts adding noise to the predictions.

\subsection{Comparison with Parallelization Strategies}

Traditional distributed training strategies maintain mathematical equivalence to single-device training through synchronized computation. Data Parallelism requires all-reduce gradient synchronization at regular intervals (every iteration for synchronous variants and every $N$ iterations for approaches like local SGD~\citep{stich2018local} and DiLoCo~\citep{douillard2023diloco}). Model Parallelism~\citep{shoeybi2020megatron} creates sequential dependencies as layers wait for gradients from subsequent stages. Pipeline Parallelism~\citep{huang2019gpipe,narayanan2019pipedream} introduces idle time during pipeline fills and drains. Each strategy imposes distinct synchronization patterns that constrain hardware deployment.

Paris's training recipe eliminates synchronization entirely. Once data partitions are established, experts train independently with no gradient synchronization, no parameter sharing, and no activation exchange. Each expert (129M-605M parameters) operates on consumer GPUs without coordination barriers or specialized interconnects. Experts train asynchronously across heterogeneous infrastructure (AWS, GCP, local clusters, Runpod) at different speeds without communication.

\begin{table}[h!]
\centering
\small
\caption{Comparison of parallelization strategies: Paris uniquely eliminates all communication overhead.}
\label{tab:parallel_comparison}
\setlength{\tabcolsep}{10pt}
\begin{tabularx}{\textwidth}{>{\raggedright\arraybackslash}p{2.8cm}>{\raggedright\arraybackslash}X>{\raggedright\arraybackslash}X>{\raggedright\arraybackslash}X}
\toprule
\textbf{Training Strategy} & \textbf{Synchronization Pattern} & \textbf{Straggler Impact} & \textbf{Topology Constraints} \\
\midrule
Data Parallel & Periodic all-reduce & Slowest worker blocks iteration & Latency-sensitive cluster \\[0.3em]
Model Parallel & Sequential layer transfers & Slowest layer blocks pipeline & Linear pipeline \\[0.3em]
Pipeline Parallel & Stage-to-stage per microbatch & Bubble overhead from slowest stage & Linear pipeline \\[0.3em]
\midrule
\textbf{Paris} & \textbf{No synchronization} & \textbf{No blocking} & \textbf{Arbitrary} \\
\bottomrule
\end{tabularx}
\end{table}

Modern optimizations including communication-computation overlap, sequence parallelism, and zero-bubble scheduling reduce synchronization overhead but cannot eliminate worker coordination requirements. Hardware heterogeneity introduces stragglers that reduce throughput proportionally, while geographic distribution amplifies latency penalties multiplicatively. Paris's zero-communication training recipe eliminates these constraints, enabling training on 120 A40 GPU-days across heterogeneous infrastructure where centralized approaches require thousands of GPU-days with specialized interconnects.

\section{Experiments and Results}
\label{sec:experiments}
\subsection{Training Details}

We conduct experiments at two model scales to validate the scalability of our decentralized diffusion framework: DiT-B/2 (129M parameters per expert) and DiT-XL/2 (605M parameters per expert), resulting in total ensemble sizes of 1.0B and 4.84B parameters respectively when accounting for 8 experts.

\subsubsection{DiT-B/2 Configuration}

The base-scale configuration employs DiT-B architecture with 768 hidden dimensions, 12 transformer layers, and 12 attention heads per expert. Training proceeds on pre-computed VAE~\citep{kingma2014auto} latents at $32 \times 32$ resolution with 4 channels, corresponding to $256 \times 256$ pixel-space images. Each expert trains independently on its assigned cluster using bath size 128 with gradient accumulation over 2 steps (effective batch 256); learning rate $1 \times 10^{-4}$ with AdamW~\citep{loshchilov2019decoupled} optimizer, no scheduling; Exponential moving average~\citep{polyak1992acceleration} with $\beta = 0.9999$ for stable inference; FP16 mixed precision training~\citep{micikevicius2018mixed} with automatic loss scaling

The corresponding router has smaller size (DiT-S) and trains separately on the full dataset with cross-entropy loss. Router training employs a smaller base batch size of 64 with 4-step gradient accumulation, yielding an effective batch of 256. We use a reduced learning rate of $5 \times 10^{-5}$ with cosine annealing~\citep{loshchilov2017sgdr} over 25 epochs to prevent overfitting on the classification task.

\subsubsection{DiT-XL/2 Configuration}

The large-scale configuration scales to DiT-XL/2 with 1152 hidden dimensions, 28 transformer layers, and 16 attention heads.

The corresponding router uses a smaller architecture (DiT-B), reducing routing overhead while maintaining classification accuracy. This asymmetric design reflects the insight that routing decisions require less capacity than generation itself.

\subsection{Results}

We evaluate generation quality using the Fréchet Inception Distance (FID)~\citep{heusel2017gans}, which measures the similarity between distributions of real and generated images in the feature space of a pretrained Inception network, with lower scores indicating better quality and diversity.

\begin{table}[t]
\centering
\small
\setlength{\tabcolsep}{10pt}
\caption{FID-50K (lower is better) comparing monolithic training and decentralized multi-expert training with different inference strategies on \textbf{Laion-art}. All models use the \textbf{DiT-B/2} architecture (129M parameters per expert).}
\label{tab:fid_mono_multi_laion}
\begin{tabular}{lc}
\toprule
\textbf{Inference Strategy} & \textbf{FID-50K $\downarrow$} \\
\midrule
Monolithic (single model)          & 29.64 \\
\midrule
Top-1                              & 30.60 \\
Top-2                              & \textbf{22.60} \\
Full Ensemble (all experts)        & 47.89 \\
\midrule
Improvement vs. Monolithic          & 7.04 \\
\bottomrule
\end{tabular}

\vspace{0.35em}
\raggedright
\textbf{Notes.} 
\textit{Monolithic}: Single model trained on full Laion-art dataset. 
\textit{Top-K}: Weighted combination of top-K experts by router confidence. 
\textit{Full}: All 8 experts weighted by router probabilities. 
Lower FID indicates better generation quality.
\end{table}

\begin{table}[t]
    \centering
    \small
    \setlength{\tabcolsep}{4pt}
    \caption{Comparison of Paris with DDM baseline on LAION-Aesthetic. Paris achieves competitive quality while using 14$\times$ fewer training images and 16$\times$ fewer GPU-days.}
    \label{tab:comparison_sota}
    \begin{tabular}{lccc@{\hskip 8pt}c}
    \toprule
    \textbf{Model} & \textbf{Params} & \textbf{Train Images} & \textbf{GPU-days} & \textbf{FID $\downarrow$} \\
    \midrule
    \textbf{Paris} (DiT-XL/2, Top-1) & 0.6B & 11M & \makecell{120 A40\\($\sim$72 A100)} & 12.45 \\
    DDM~\citep{mcallister2025decentralized} & 0.89B & 154M & \makecell{168 GPUs $\times$ 7d\\($\sim$1176 A100)} & 5.5--10.5$^\dagger$ \\
    \midrule
    \rowcolor{orange!20}
    \textbf{Ratio (DDM/Paris)} & 1.48$\times$ & 14.0$\times$ & 16.3$\times$ & 0.44--0.84$\times$ \\
    \bottomrule
    \end{tabular}
    
    \vspace{0.35em}
    \raggedright
    \footnotesize
    \textbf{Notes:} Paris adopts DiT-XL/2 with architectural optimizations from PixArt-$\alpha$, reducing parameters from 0.89B to 0.6B. $^\dagger$~\citet{mcallister2025decentralized} does not report a single FID on LAION-Aesthetic; the range 5.5--10.5 is estimated from results at varying training FLOPs in their paper.
\end{table}

\begin{figure*}[t]
    \centering
    \includegraphics[width=1\linewidth]{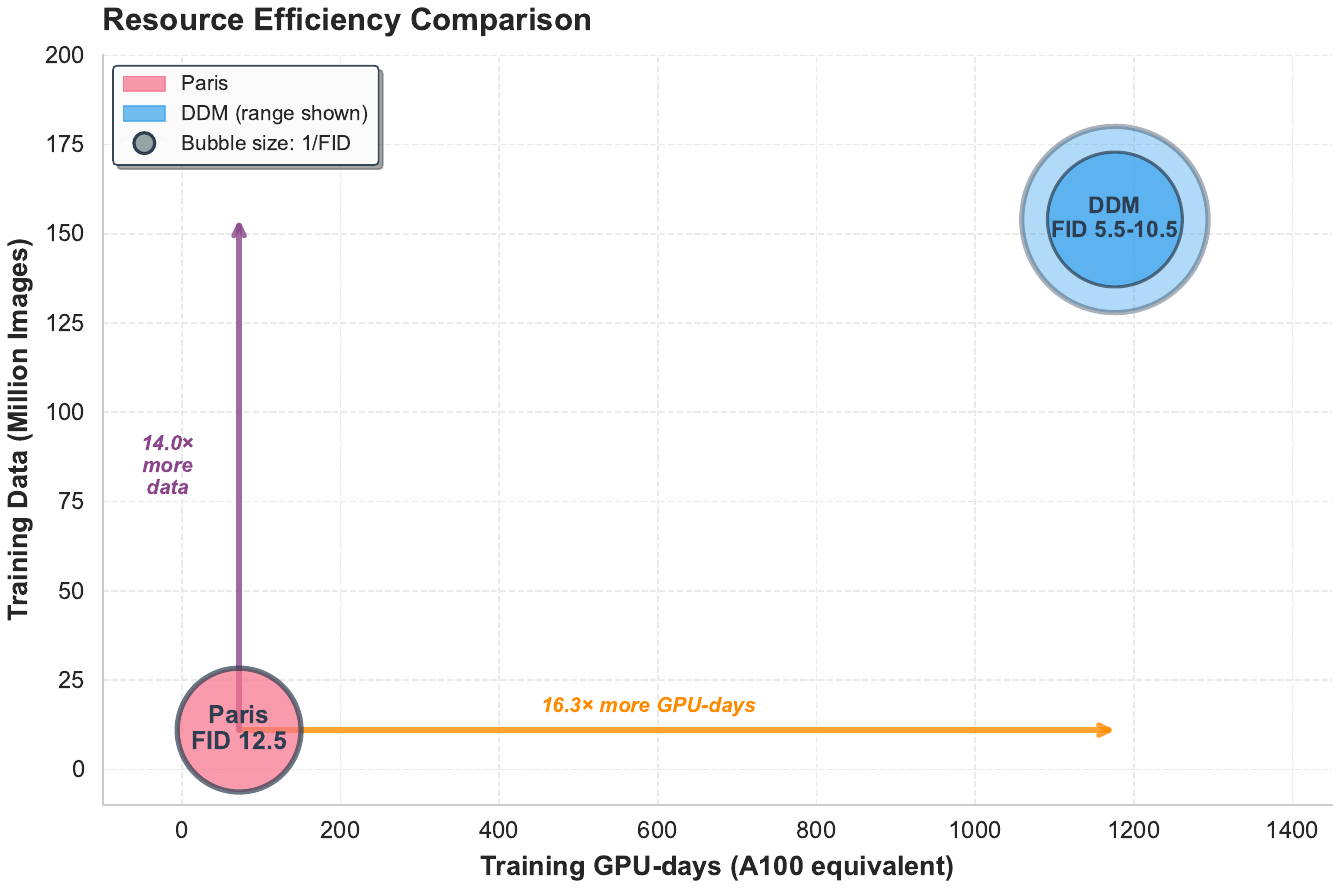}
        \caption{Efficiency comparison of Paris and DDM.}
    \label{fig:efficiency_plot}
\end{figure*}

    Table~\ref{tab:fid_mono_multi_laion} compares inference strategies for our multi-expert approach against a monolithic baseline, both using the DiT-B/2 architecture (129M parameters per expert). The Top-2 strategy, which uses a weighted combination of the top-2 experts by router confidence, achieves the best performance with FID-50K of 22.60, representing a 7.04 improvement over the monolithic baseline (29.64). Surprisingly, the Full Ensemble strategy that weights all 8 experts by router probabilities underperforms (47.89), suggesting that selective expert collaboration is more effective than naive averaging across all experts. This validates our hypothesis that targeted routing outperforms simple ensemble approaches with Paris.

Table~\ref{tab:comparison_sota} and Figure~\ref{fig:efficiency_plot} compare Paris with DDM~\citep{mcallister2025decentralized}. Paris adopts DiT-XL/2 with architectural optimizations from PixArt-$\alpha$, reducing parameters from 0.89B to 0.6B. Paris achieves FID of 12.45 on LAION-Aesthetic using 11M training images and 120 A40 GPU-days ($\sim$72 A100 equivalent), while DDM achieves FID in the range 5.5--10.5 (depending on training FLOPs) using 154M images and $\sim$1176 A100 GPU-days. Paris uses 14$\times$ less training data and 16$\times$ fewer GPU-days, with 1.19--2.26$\times$ higher FID. These results demonstrate that decentralized diffusion training can achieve competitive generation quality with substantially improved resource efficiency, making high-quality text-to-image generation accessible with commodity hardware and limited data.

\section{Conclusion}

We demonstrate that high-quality diffusion models can be trained through fully decentralized computation, achieving competitive text-to-image synthesis without gradient synchronization. This work establishes both the feasibility and practical implementation patterns for distributed generative modeling, opening pathways toward larger-scale decentralized training. Future work includes model distillation for deployment efficiency, adaptation to video generation, and incorporation of architectural advances.

\bibliography{papers}
\bibliographystyle{plainnat}

\end{document}